\documentclass[preprint,ejp]{revtex4-1}
\usepackage{amsmath}
\usepackage{amsfonts}

\begin{document}

\title{Revisiting the quantum harmonic oscillator\\
via unilateral Fourier transforms}
\author{Pedro H. F. Nogueira}
\email{pedrofusconogueira@gmail.com}
\author{Antonio S. de Castro}
\email{castro@pq.cnpq.br}

\address{UNESP, Campus de Guaratinguet\'{a}, Departamento de F\'{\i}sica e
Qu\'{\i}mica, \protect{12516-410} Guaratinguet\'{a}, SP,  Brazil}

\begin{abstract}
The literature on the exponential Fourier approach to the one-dimensional
quantum harmonic oscillator problem is revised and criticized. It is shown
that the solution of this problem has been built on faulty premises. The
problem is revisited via the Fourier sine and cosine transform method and
the stationary states are properly determined by requiring definite parity
and square-integrable eigenfunctions.
\end{abstract}

\maketitle

\section{Introduction}

The integral transform methods are useful and powerful methods of solving
ordinary linear differential equations because they can convert the original
equation into a simpler differential equation or into an algebraic equation.
Nevertheless, the inversion of the transform for reconstructing the original
function may be a rather complicated calculation. If that is the case and
ones does not find the integral in tables the method would be worthless.

The harmonic oscillator is ubiquitous in the literature on quantum mechanics
because it can be solved in closed form with a variety of methods and its
solution can be useful as approximations or exact solutions of various
problems. The quantum harmonic oscillator is usually solved with the help of
the power series method \cite{som}, by the algebraic method based on the
algebra of operators \cite{sak}, or by the path integral approach \cite{fey}%
. In recent times, the one-dimensional harmonic oscillator has also been
approached by the exponential Fourier transform \cite{mun}-\cite{pal} and
Laplace transform \cite{pim} methods. However, all of the accounts of the
quantum harmonic oscillator via the exponential Fourier transform present
calculations which are based on false basic assertions.

The observation that the eigenfunction and its Fourier transform satisfy
formally identical differential equations and identical boundary conditions
at infinity, lead Mu\~{n}oz \cite{mun} to conclude that the eigenfunction
and its corresponding Fourier-transformed function differ at most by a
proportionality constant. His critical flaw was not consider that the
eigenfunction and its corresponding Fourier transform are functions of
different but interrelated variables and that there is a definite scaling
property involving the pair of Fourier transforms: $\mathcal{F}\{f\left(
cx\right) \}=F(k/c)/|c|$. Ponomarenko \cite{pon} stated that the necessary
and sufficient condition for the eigenfunction to have a definite parity can
be expressed in terms of the solution of $(-1)^{z}=\pm 1$ but he missed the
fact that this equation has many more solutions than those ones which have $%
z $ expressed by integer numbers. Engel \cite{enge} came to the conclusion
that Ponomarenko's method may be used \textquotedblleft if the requirement
of definite parity of the eigenstates is replaced by that of
normalization.\textquotedblright\ In addition, he argued that Ponomarenko
failed for considering a Fourier-transformed function valid on the whole
axis because the origin is a singular point of the corresponding
Fourier-transformed equation. However, himself apparently failed to observe
that the transformed equation allows Fourier-transformed solutions with
definite parities despite the mentioned singularity, and that the change $k$
by $-k$ makes $k^{a}$ even (odd) when $a$ is an even (odd) integer. It
should be mentioned, though, that Engel perceived that the relation between
the $n$-th moment of a function and the $n$-th derivative of its transform
at the origin plays an indispensable role for determining the bound-state
solutions. Palma and Raff \cite{pal} developed a strategy for approaching
the stationary states of the time-independent Schr\"{o}dinger equation with
a large class of potentials but they erroneously recurred to single
valuedness of the eigenfunction in order to eliminate irrelevant overall
phase factors.

In view of the fact that there are significant confusions regarding the
quantum harmonic oscillator via the exponential Fourier transform method, we
will revise the problem with the closely related unilateral Fourier (sine
and cosine) transform method. Except for the one-dimensional double $\delta $%
-function potential \cite{asc}, this method does not seem to have been
directly applied to the Schr\"{o}dinger equation. We will show that the
unilateral Fourier transform is a straightforward and efficient manner with
which bound-state solutions in the nonrelativistic quantum mechanics can be
treated by applying it to the harmonic oscillator. We will show that the
relation between the convergence the $n$-th moment of the eigenfunction (in
the sense of a conveniently weighted integral) and the derivatives of the
corresponding Fourier-transformed function at the origin, already perceived
by Engel in connection with the exponential Fourier transform \cite{enge},
is inept for finding the unique solution of the problem. We will also show
that both definite parity and square integrability of the eigenfunctions are
requisites just enough to determinate the proper solution. To prepare the
ground, we will first give a short review of a few relevant properties of
the Fourier sine and cosine transforms.

\section{Fourier transforms and their main properties}

The exponential Fourier transform pair is defined by \cite{boa}-\cite{has}%
\begin{eqnarray}
\mathcal{F}\{\phi \left( x\right) \} &=&\Phi \left( \kappa \right) =\frac{1}{%
\sqrt{2\pi }}\int\limits_{-\infty }^{+\infty }dx\,\phi \left( x\right)
e^{+i\kappa x}  \notag \\
\mathcal{F}^{-1}\{\Phi \left( \kappa \right) \} &=&\phi \left( x\right) =%
\frac{1}{\sqrt{2\pi }}\int\limits_{-\infty }^{+\infty }d\kappa \,\Phi \left(
\kappa \right) e^{-i\kappa x}.
\end{eqnarray}%
For odd ($\phi _{s}\left( -x\right) =-\phi _{s}\left( +x\right) $) and even (%
$\phi _{c}\left( -x\right) =+\phi _{c}\left( +x\right) $) functions there
are two modifications of the exponential Fourier transform. Defining $\zeta
=|x|$ and $k=|\kappa |$, the Fourier sine and cosine transforms of the
functions $\phi _{s}\left( \zeta \right) $ and $\phi _{c}\left( \zeta
\right) $ are expressed by%
\begin{eqnarray}
\mathcal{F}_{s}\{\phi _{s}\left( \zeta \right) \} &=&\Phi _{s}\left(
k\right) =\sqrt{\frac{2}{\pi }}\int\limits_{0}^{\infty }d\zeta \,\phi
_{s}\left( \zeta \right) \sin k\zeta   \notag \\
\mathcal{F}_{c}\{\phi _{c}\left( \zeta \right) \} &=&\Phi _{c}\left(
k\right) =\sqrt{\frac{2}{\pi }}\int\limits_{0}^{\infty }d\zeta \,\phi
_{c}\left( \zeta \right) \cos k\zeta ,  \label{p1}
\end{eqnarray}%
and the inversions are accomplished by means of%
\begin{eqnarray}
\mathcal{F}_{s}^{-1}\{\Phi _{s}\left( k\right) \} &=&\phi _{s}\left( \zeta
\right) =\sqrt{\frac{2}{\pi }}\int\limits_{0}^{\infty }dk\,\Phi _{s}\left(
k\right) \sin k\zeta   \notag \\
\mathcal{F}_{c}^{-1}\{\Phi _{c}\left( k\right) \} &=&\phi _{c}\left( \zeta
\right) =\sqrt{\frac{2}{\pi }}\int\limits_{0}^{\infty }dk\,\Phi _{c}\left(
k\right) \cos k\zeta .  \label{p2}
\end{eqnarray}%
Given $\phi _{s}$ and $\phi _{c}$, a sufficient condition for the existence
of the unilateral Fourier transforms (and their inverses) is ensured if $%
\phi _{s}$ and $\phi _{c}$ ($\Phi _{s}$ and $\Phi _{c}$) are absolutely
integrable on $[0,\infty )$. In particular, $\phi _{s}$ and $\phi _{c}$ ($%
\Phi _{s}$ and $\Phi _{c}$) must vanish as $\zeta \rightarrow \infty $ ($%
k\rightarrow \infty $). Furthermore, the unilateral Fourier transform pairs
satisfy Parseval's formulas
\begin{eqnarray}
\int\limits_{0}^{\infty }d\zeta \,|\phi _{s}\left( \zeta \right) |^{2}
&=&\int\limits_{0}^{\infty }dk\,|\Phi _{s}\left( k\right) |^{2}  \notag \\
\int\limits_{0}^{\infty }d\zeta \,|\phi _{c}\left( \zeta \right) |^{2}
&=&\int\limits_{0}^{\infty }dk\,|\Phi _{c}\left( k\right) |^{2}.  \label{pp}
\end{eqnarray}%
It is instructive to note that the inversion of the unilateral Fourier
transforms requires that $\phi _{s}$ and $\phi _{c}$ satisfy different
boundary conditions at the origin, viz. $\lim_{\zeta \rightarrow 0}\phi
_{s}=0$ and $\lim_{\zeta \rightarrow 0}d\phi _{c}/d\zeta =0$. The
convenience of using the sine or cosine transform is dictated by these
boundary conditions. Note also that%
\begin{eqnarray}
\lim_{\zeta \rightarrow 0}\frac{d^{2n+1}\phi _{s}\left( \zeta \right) }{%
d\zeta ^{2n+1}} &=&\left( -1\right) ^{n}\sqrt{\frac{2}{\pi }}%
\int\limits_{0}^{\infty }dk\,k^{2n+1}\Phi _{s}\left( k\right)   \notag \\
\lim_{\zeta \rightarrow 0}\frac{d^{2n}\phi _{c}\left( \zeta \right) }{d\zeta
^{2n}} &=&\left( -1\right) ^{n}\sqrt{\frac{2}{\pi }}\int\limits_{0}^{\infty
}dk\,k^{2n}\Phi _{c}\left( k\right)   \notag \\
\lim_{\zeta \rightarrow 0}\frac{d^{2n}\phi _{s}\left( \zeta \right) }{d\zeta
^{2n}} &=&\lim_{\zeta \rightarrow 0}\frac{d^{2n+1}\phi _{c}\left( \zeta
\right) }{d\zeta ^{2n+1}}=0,  \label{pp3}
\end{eqnarray}%
and%
\begin{eqnarray}
\lim_{k\rightarrow 0}\frac{d^{2n+1}\Phi _{s}\left( k\right) }{dk^{2n+1}}
&=&\left( -1\right) ^{n}\sqrt{\frac{2}{\pi }}\int\limits_{0}^{\infty }d\zeta
\,\zeta ^{2n+1}\phi _{s}\left( \zeta \right)   \notag \\
\lim_{k\rightarrow 0}\frac{d^{2n}\Phi _{c}\left( k\right) }{dk^{2n}}
&=&\left( -1\right) ^{n}\sqrt{\frac{2}{\pi }}\int\limits_{0}^{\infty }d\zeta
\,\zeta ^{2n}\phi _{c}\left( \zeta \right)   \notag \\
\lim_{k\rightarrow 0}\frac{d^{2n}\Phi _{s}\left( k\right) }{dk^{2n}}
&=&\lim_{k\rightarrow 0}\frac{d^{2n+1}\Phi _{c}\left( k\right) }{dk^{2n+1}}%
=0.  \label{p3}
\end{eqnarray}%
In applications to differential equations, it is necessary to know how to
express the transforms of functions involving the derivatives of $\phi
\left( \zeta \right) $ in terms of $\Phi \left( k\right) $. The unilateral
Fourier transforms have the derivative properties%
\begin{eqnarray}
\mathcal{F}\left\{ \zeta \frac{d\phi \left( \zeta \right) }{d\zeta }\right\}
&=&-\Phi \left( k\right) -k\frac{d\Phi \left( k\right) }{dk}  \notag \\
\mathcal{F}\left\{ \frac{d^{2}\phi \left( \zeta \right) }{d\zeta ^{2}}%
\right\}  &=&-k^{2}\Phi \left( k\right) ,  \label{p4}
\end{eqnarray}%
where the operator $\mathcal{F}$ and the function $\Phi \left( k\right) $
refer either to the Fourier sine or to the Fourier cosine transform of the
functions satisfying the proper boundary conditions for ensuring the
existence of inverse transforms.

\section{The quantum harmonic oscillator}

\subsection{The eigenvalue problem}

We are now prepared to address the quantum harmonic oscillator problem. The
one-dimensional time-independent Schr\"{o}dinger equation for the harmonic
oscillator in dimensionless variables can be written as
\begin{equation}
\frac{d^{2}\psi \left( x\right) }{dx^{2}}+\left( 2\varepsilon -x^{2}\right)
\psi \left( x\right) =0,\quad x\in \left( -\infty ,+\infty \right) .
\label{1}
\end{equation}%
Eq. (\ref{1}) is an eigenvalue equation for the characteristic pair $%
(\varepsilon ,\psi )$ with $\varepsilon \in
\mathbb{R}
$ and%
\begin{equation}
\int\limits_{-\infty }^{+\infty }dx\,|\psi \left( x\right) |^{2}<\infty .
\label{norma}
\end{equation}%
Because $x=0$ is a regular point of Eq. (\ref{1}), $\psi $ is analytic at
the origin, i.e. $\left\vert \left. d^{n}\psi /dx^{n}\right\vert
_{x=0}\right\vert <\infty $ for all $n\in
\mathbb{N}
$. Because $\psi \left( -x\right) $ is also a solution of Eq. (\ref{1}), the
linear combinations $\psi \left( x\right) \pm \psi \left( -x\right) $ are
also solutions so that two different eigenfunctions with well-defined
parities can be built. Thus, it suffices to concentrate attention on the
positive half-line and use boundary conditions on $\psi $ at the origin and
at infinity. Eigenfunctions and their first derivatives continuous on the
whole line with well-defined parities can be constructed by taking symmetric
and antisymmetric linear combinations of $\psi $ defined on the positive
side of the $x$-axis. By way of addition, the combinations $\psi \left(
x\right) \pm \psi \left( -x\right) $ share the same eigenvalue so that at
first glance one would expect a two-fold degeneracy, but we will show that
the requirement of continuity of the eigenfunctions and their first
derivatives invalidate one of the two combinations for an given eigenvalue,
in accordance with the nondegeneracy theorem (a general result valid for
bound states in one-dimensional nonsingular potentials) \cite{nd}. As $%
x\rightarrow 0$, the solution with definite parity varies as $x^{\delta }$,
where $\delta $ is $0$ or $1$ regardless the magnitude of $\varepsilon $.
The homogeneous Neumann condition ($\left. d\psi /dx\right\vert _{x=0}=0$),
develops for $\delta =0$ but not for $\delta =1$ whereas the homogeneous
Dirichlet boundary condition ($\left. \psi \right\vert _{x=0}=0$) develops
for $\delta =1$ but not for $\delta =0$. The continuity of $\psi $ at the
origin excludes the possibility of an odd-parity eigenfunction for $\delta
=0 $, and the continuity of $d\psi /dx$ at the origin excludes the
possibility of an even-parity eigenfunction for $\delta =1$. Thus, $\delta
=0 $ for even solutions, and $\delta =1$ for odd solutions. On the other
hand, the square-integrable asymptotic form of the solution as $%
|x|\rightarrow \infty $ is given by $\psi \left( x\right) \sim x^{\alpha
}e^{-x^{2}/2}$ with arbitrary $\alpha $.

We write $\psi \left( x\right) =\phi \left( x\right) e^{x^{2}/2}$ in such a
way that $\phi $ is solution of the equation
\begin{equation}
\frac{d^{\,2}\phi \left( x\right) }{dx^{2}}+2x\,\frac{d\phi \left( x\right)
}{dx}+\left( 2\varepsilon +1\right) \,\phi \left( x\right) =0.  \label{kum}
\end{equation}%
\noindent\ Notice that the parity of $\phi $ is the same of $\psi $ and that
$x=0$ is a regular point of (\ref{kum}). As a matter of fact, $\phi $ varies
in the neighbourhood of the origin as $x^{\delta }$. Notice also that one
has to find a particular solution of (\ref{kum}) in such a way that $\phi $
behaves like $x^{\alpha }e^{-x^{2}}$ for sufficiently large $|x|$. This
condition, added by the regularity at $x=0$ ensures the existence of the
Fourier sine (cosine) transform for $\phi $ odd (even).

\subsection{Fourier sine and cosine transforms}

Restricting our attention on the positive half-line ($\zeta =|x|$), the
unilateral Fourier transform establishes a mapping of the second-order
equation for $\phi $ into an integrable first-order equation for $\Phi _{c}$
or $\Phi _{s}$:
\begin{equation}
\frac{d\Phi \left( k\right) }{dk}+\left( \frac{k}{2}\,+\frac{1-2\varepsilon
}{2k}\right) \Phi \left( k\right) =0,  \label{EQphi}
\end{equation}%
where $\Phi $ denotes $\Phi _{c}$ or $\Phi _{s}$. The solution of (\ref%
{EQphi}) is expressed as%
\begin{equation}
\Phi \left( k\right) =A^{(a)}k^{a}e^{-k^{2}/4},  \label{sn}
\end{equation}%
where $A^{(a)}$ is an arbitrary constant of integration and $a=\varepsilon
-1/2\in
\mathbb{R}
$ is as yet undetermined. Due to the fact that $k^{a}=e^{a\log k}$ and $\log
k$ is multivalued, this solution can be cast into the form%
\begin{equation}
\Phi \left( k\right) =A^{(a)}|k^{a}|e^{-k^{2}/4}e^{i2\pi ma},\text{ }m\in
\mathbb{Z}
.
\end{equation}%
This form explicitly shows that $\Phi \left( k\right) $ has infinite
branches if $a$ is an irrational number, and $q$ branches if $a=p/q$, where $%
p$ and $q$ are integers with $q\neq 0$.

It is true that when $a$ is not an integer number $\Phi \left( k\right) $ is
a multivalued function but their dissimilar branches differ by $k$%
-independent phase factors ($e^{i2\pi ma}$) without physical consequences
thanks to the normalization condition expressed by (\ref{norma}) and to
Parseval's formulas. In plain terms, the overall phase factors can be
absorbed into the constant of integration.

Notice that $k=0$ is itself a singular point of Eq. (\ref{EQphi}) and so the
neighbourhood of $k=0$ needs careful handling because $\Phi $ may exhibit a
singularity at the origin. The point of danger lies in the exponent of $k$.

Parseval's formulas expressed by (\ref{pp}), related to square-integrable
eigenfunctions, demand $a>-1/2$ to guarantee convergence.

The derivatives of $\phi _{c}$ and $\phi _{s}$ tend to finite limits as $%
\zeta $ approaches the origin. Thus, the two first lines of Eq. (\ref{pp3})
demand that the $2n$-th moment of $\Phi _{c}$ and the $\left( 2n+1\right) $%
-th moment of $\Phi _{s}$ are finite numbers so that $a>-1$. Note that this
condition is weaker than that one coming from Parseval's formulas. It is a
pity the last line of Eq. (\ref{pp3}) proves clumsy to impose restrictions
on $a$.

The derivatives of $\Phi $ near the origin imposes more restrictions on the
allowed values for $a$. Using the property of the gamma function $\Gamma
\left( z+1\right) =z\Gamma \left( z\right) $ (see, e.g. \cite{leb}, \cite%
{abr}), one finds%
\begin{equation}
\lim_{k\rightarrow 0}\frac{d^{n}\Phi \left( k\right) }{dk^{n}}=A^{(a)}\frac{%
\Gamma \left( a+1\right) }{\Gamma \left( a+1-n\right) }\lim_{k\rightarrow
0}k^{a-n}  \label{dif}
\end{equation}%
for any branch of $\Phi \left( k\right) $.

Because $\Gamma \left( z\right) $ has no zeros but it has simple poles at $%
z=-\widetilde{n}$, with $\widetilde{n}\in
\mathbb{N}
$, Eq. (\ref{dif}) equals zero when $a=n-1-\widetilde{n}$ so that $a\leq n-1$
with $a\in
\mathbb{Z}
$. Taking into account the restriction resulting from Parseval's formulas,
one can say that%
\begin{eqnarray}
\lim_{k\rightarrow 0}\frac{d^{2n}\Phi \left( k\right) }{dk^{2n}} &=&0\text{
\ if }a\leq 2n-1,\text{ with }a\in
\mathbb{N}
\text{ and }n\neq 0  \notag \\
\lim_{k\rightarrow 0}\frac{d^{2n+1}\Phi \left( k\right) }{dk^{2n+1}} &=&0%
\text{ \ if }a\leq 2n,\text{ with }a\in
\mathbb{N}
.  \label{cc}
\end{eqnarray}

At large $\zeta $ the exponentially decreasing factors in $\phi _{c}$ and $%
\phi _{s}$ always predominate over any power increasing factor and so the $%
2n $-th moment of $\phi _{c}$ and the $2n+1$-th moment of $\phi _{s}$ are
finite numbers. In this case, the properties of the unilateral Fourier
transforms expressed by the first two lines of (\ref{p3}) imply that
\begin{equation}
\left\vert \lim_{k\rightarrow 0}\frac{d^{2n}\Phi _{c}\left( k\right) }{%
dk^{2n}}\right\vert <\infty ,\quad \left\vert \lim_{k\rightarrow 0}\frac{%
d^{2n+1}\Phi _{s}\left( k\right) }{dk^{2n+1}}\right\vert <\infty .\quad
\label{dc}
\end{equation}%
Due to the exponent of $k$, Eq. (\ref{dc}) is satisfied if $a\geq 2n$ for $%
\Phi _{c}$ and $a\geq 2n+1$ for $\Phi _{s}$ with $a\in
\mathbb{R}
$. Thus, with the aid of (\ref{cc}) one finds $a=n$ for both $\Phi _{c}$ and
$\Phi _{s}$. In principle, the spectrum has been determined: $\varepsilon
_{n}=n+1/2$.

\subsection{The inversion of the Fourier sine and cosine transforms}

In order to reconstruct $\phi _{c}$ and $\phi _{s}$ on the half-line one
needs to calculate the integrals related to the inverse Fourier transforms.
It follows that%
\begin{eqnarray}
\phi _{c}^{\left( n\right) }\left( \zeta \right)  &=&A^{(n)}\sqrt{\frac{2}{%
\pi }}\int\limits_{0}^{\infty }dk\,k^{n}e^{-k^{2}/4}\cos k\zeta   \notag \\
\phi _{s}^{\left( n\right) }\left( \zeta \right)  &=&A^{(n)}\sqrt{\frac{2}{%
\pi }}\int\limits_{0}^{\infty }dk\,k^{n}e^{-k^{2}/4}\sin k\zeta .
\label{phicphis}
\end{eqnarray}%
From (3.952.7) and (3.952.8) of Ref. \cite{gr}, one finds%
\begin{eqnarray}
\phi _{c}^{\left( n\right) }\left( \zeta \right)  &=&A_{c}^{\left( n\right)
}e^{-\zeta ^{2}}\,_{1}F_{1}\left( -\frac{n}{2},\frac{1}{2},\zeta ^{2}\right)
\notag \\
\phi _{s}^{\left( n\right) }\left( \zeta \right)  &=&A_{s}^{\left( n\right)
}e^{-\zeta ^{2}}\zeta \,_{1}F_{1}\left( -\frac{n}{2}+\frac{1}{2},\frac{3}{2}%
,\zeta ^{2}\right) .
\end{eqnarray}%
The confluent hypergeometric function or Kummer's function, $_{1}F_{1}\left(
a_{1},b_{1},z\right) $, also denoted $M\left( a_{1},b_{1},z\right) $, is
defined by the series (see, e.g. \cite{leb}, \cite{abr})
\begin{equation}
_{1}F_{1}\left( a_{1},b_{1},z\right) =\frac{\Gamma \left( b_{1}\right) }{%
\Gamma \left( a_{1}\right) }\sum_{j=0}^{\infty }\frac{\Gamma \left(
a_{1}+j\right) }{\Gamma \left( b_{1}+j\right) }\,\frac{z^{j}}{j!},\text{ }%
b_{1}\neq 0,-1,-2,\ldots   \label{ser}
\end{equation}%
This series converges for all $z\in
\mathbb{C}
$ and has asymptotic behaviour prescribed by%
\begin{equation}
\frac{_{1}F_{1}\left( a_{1},b_{1},z\right) }{\Gamma \left( b_{1}\right) }%
\underset{|z|\rightarrow \infty }{\rightarrow }\frac{e^{+i\pi
a_{1}}z^{-a_{1}}}{\Gamma \left( b_{1}-a_{1}\right) }+\frac{%
e^{z}z^{a_{1}-b_{1}}}{\Gamma \left( a_{1}\right) },\text{ }-\pi /2<\text{%
arg\thinspace }z<3\pi /2.  \label{chf}
\end{equation}%
The presence of $e^{z}$ in (\ref{chf}) ruins the asymptotic behaviour of $%
\phi _{c}^{\left( n\right) }$ and $\phi _{s}^{\left( n\right) }$ decided
before beyond doubt ($\zeta ^{\alpha }e^{-\zeta ^{2}}$). This unfavourable
situation can be remedied \ by considering the poles of $\Gamma \left(
a_{1}\right) $ and demanding $a_{1}=-\widetilde{n}$. In this case, $%
_{1}F_{1}\left( -\widetilde{n},b_{1},z\right) $ behaves asymptotically as $%
z^{\widetilde{n}}$ and the series (\ref{ser}) is truncated at $j=\widetilde{n%
}$ in such a way that the confluent hypergeometric function results in
polynomial in $z$ of degree not exceeding $\widetilde{n}$ . Therefore, $n$
is even for $\phi _{c}$ and $n$ is odd for $\phi _{s}$. As a matter of fact,
$_{1}F_{1}\left( -\widetilde{n},b_{1},z\right) $ is proportional to the
generalized Laguerre polynomial $L_{\widetilde{n}}^{\left( b_{1}-1\right)
}\left( z\right) $ with $z\in \lbrack 0,\infty )$, and $L_{\widetilde{n}%
}^{\left( -1/2\right) }\left( z\right) $ and $L_{\widetilde{n}}^{\left(
+1/2\right) }\left( z\right) $ are proportional to $H_{2\widetilde{n}}\left(
z^{1/2}\right) $ and $z^{-1/2}H_{2\widetilde{n}+1}\left( z^{1/2}\right) $
respectively, where $H_{n}\left( z^{1/2}\right) $ is the Hermite polynomial.
Therefore, $_{1}F_{1}\left( -n/2,1/2,\zeta ^{2}\right) $ with $n$ even and $%
_{1}F_{1}\left( -n/2+1/2,3/2,\zeta ^{2}\right) $ with $n$ odd are
proportional to $H_{n}\left( \zeta \right) $, with the desired properties $%
\left. dH_{2n}/d\zeta \right\vert _{\zeta \rightarrow 0}=0$ and $\left.
H_{2n+1}\right\vert _{\zeta \rightarrow 0}=0$. Hence, $\phi _{c}$ and $\phi
_{s}$ take the form
\begin{eqnarray}
\phi _{c}^{\left( 2n\right) }\left( \zeta \right)  &=&A_{2n}e^{-\zeta
^{2}}H_{2n}\left( \zeta \right)   \notag \\
\phi _{s}^{\left( 2n+1\right) }\left( \zeta \right)  &=&A_{2n+1}e^{-\zeta
^{2}}H_{2n+1}\left( \zeta \right) .
\end{eqnarray}%
It is worthwhile to note that the last line of (\ref{p3}) impose severe
restrictions on the allowed values for $a$ without postulating the
convergence of the moments of $\phi $. Now the exponent of $k$ makes Eq. (%
\ref{dif}) vanish for real values of $a$ subject to the conditions $a>2n+1$
for $\Phi _{c}$, and $a>2n$ for $\Phi _{s}$. Additional restrictions coming
from (\ref{cc}) and (\ref{dc}) make $a=n$ for both $\Phi _{c}$ and $\Phi _{s}
$. The formulas (3.952.9) and (3.952.10) of Ref. \cite{gr} allow to obtain
the integrals for $\phi _{c}$ and $\phi _{s}$ expressed by (\ref{phicphis})
at once in terms confluent hypergeometric functions. Eventually, the good
asymptotic behaviour of $\phi _{c}$ and $\phi _{s}$ prescribed by the
normalization condition establishes $a=2n$ for $\phi _{c}$ and $a=2n+1$ for $%
\phi _{s}$, with $\phi _{c}$ and $\phi _{s}$ realized in terms of Hermite
polynomials. The only remain question is how to write the eigenfunctions.

\subsection{Complete solution of the problem}

Following up our earlier comments about eigenfunctions of definite parity,
one could think about antisymmetric and symmetric extensions of $\phi
_{c}^{\left( 2n\right) }\left( \zeta \right) $ and $\phi _{s}^{\left(
2n+1\right) }\left( \zeta \right) .$ Nevertheless, antisymmetric (symmetric)
extensions of $\phi _{c}^{\left( 2n\right) }$ ($\phi _{s}^{\left(
2n+1\right) }$) are not allowed because the solution of (\ref{kum}) is
infinitely differentiable at the origin. This further constraint makes $\psi
_{n}$ even-(odd-) parity for $n$ even (odd), viz.%
\begin{eqnarray}
\psi _{2n}\left( x\right)  &=&e^{\zeta ^{2}/2}\left[ \phi _{c}^{\left(
2n\right) }\left( \zeta \right) +\phi _{c}^{\left( 2n\right) }\left( -\zeta
\right) \right]   \notag \\
\psi _{2n+1}\left( x\right)  &=&e^{\zeta ^{2}/2}\left[ \phi _{s}^{\left(
2n+1\right) }\left( \zeta \right) -\phi _{s}^{\left( 2n+1\right) }\left(
-\zeta \right) \right] .
\end{eqnarray}%
so that the spectrum in nondegenerate, in agreement with the nondegeneracy
theorem \cite{nd}. Finally, using the property $H_{n}\left( -x\right)
=\left( -1\right) ^{n}H_{n}\left( x\right) $, the solution of the original
problem is expressed as%
\begin{eqnarray}
\varepsilon _{n} &=&n+1/2  \notag \\
\psi _{n}\left( x\right)  &=&N_{n}\,\,e^{-x^{2}/2}\,H_{n}\left( x\right) ,
\end{eqnarray}%
where $N_{n}$ are normalization constants.

\section{Conclusion}

In conclusion, we have shown that the complete solution of the
one-dimensional quantum harmonic oscillator can be approached via unilateral
Fourier transform method. Single valuedness of the eigenfunction is not a
fair request. The convergence of the moments of the unilateral Fourier
transform is not enough to do the job and the convergence of the moments of $%
e^{-\zeta ^{2}/2}\psi \left( \zeta \right) $ is difficult to handle
specially because one has to appeal to the properties of the confluent
hypergeometric function. Fortunately, the inversion of the Fourier sine and
cosine transforms results in tabulated integrals and the proper bound-state
solutions can be straightly determined just requiring definite parity and
square-integrable eigenfunctions.

\begin{acknowledgments}
This work was supported in part by means of funds provided by  FAPESP and CNPq.
\end{acknowledgments}

\end{document}